\documentstyle[12pt,aasms]{article}

\begin{document}

\title{Giant Branch Mixing and the Ultimate Fate of Primordial Deuterium
in the Galaxy}
\author{Craig J. Hogan}
\affil{Astronomy Department FM-20, University of Washington,
    Seattle, WA 98195}
\def\msol{{\,\rm M_{\odot}}}
\begin{abstract}
The  observed cosmic abundances of light elements are most
consistent with
each other, and with the predictions of big bang nucleosynthesis,
if, contrary to the usual assumption,
 galactic chemical evolution reduces $(D+\ ^3He)/H$ with time.
Chemical evolution models which do this
 require that low mass stars  destroy $^3He$ in the
envelope gas that they   return  to the interstellar medium.
A  simple argument based on the rates of limiting nuclear reactions
shows that the  same giant branch mixing process which appears to be needed to
explain the observed  $^{12}C/ ^{13}C$ and $C/N$ ratios in 1-- 2$\msol$
stars would
indeed also probably destroy $^3He$ by a large factor in the bulk of the
envelope material.
The conclusion is that
Galactic $^3He/H$ estimates should not be trusted for setting an upper
limit on
 primordial $(D+  ^3He)/H$. This removes the   strongest lower bound
on the cosmic baryon density from big bang nucleosynthesis and the only
argument for
abundant baryonic dark matter.
\end{abstract}


\section{Is $D+^3He$ Destroyed in the Galaxy?}

Almost all cosmic deuterium is a relic of primordial
nucleosynthesis. But since the big bang, successive generations of
stars have destroyed most  of it by burning to heavier
elements. Constraints on the primordial deuterium
abundance $(D/H)_p$  have been deduced from $D$ abundances in the
reprocessed material in the Galactic interstellar medium and from
 the local abundances of the principal immediate product
of $D$ burning, $^3He$.
I suggest here  that these deductions may be wrong, because
the $^3He$ abundance in the Galaxy
is   likely to be strongly affected by the post main sequence mixing
that seems to be required by   observations of
 $^{12}C/ ^{13}C$ and $C/N$ in low mass stars.
If this is so,  Galactic $^3He/H$  estimates should not be trusted
as indicators of the primordial $D/H$    abundance,
because they are so sensitive to uncalibrated   aspects of
stellar evolution models. This argument suggests that
  primordial $D/H$ should  be estimated  instead from
    relatively unprocessed
material, such as
 protogalactic quasar absorbers at high redshift.

It is important to resolve this question. The   abundance of
 $^3He$  currently gives the only upper limit
on the primordial $D/H$,  and thereby  the only strong lower limit on
the cosmic baryon-to-photon ratio $\eta$ from standard big bang
nucleosynthesis theory (SBBN; see eg Smith, Kawano, and Malaney 1993,
Copi, Schramm and Turner 1994)---
in particular, the only lower limit appreciably  higher
than observed density of baryons. It therefore
provides the only argument for abundant baryonic dark matter.
If this lower limit were relaxed, the range of allowed
 $\eta$ includes lower values where the predicted $^4He$ abundance lies
more comfortably close to observations,  where the bound on the
number of particle species is  relaxed, and where there is almost no
baryonic dark matter.

In the standard picture    the bulk
of   primordial $D$  in the Galaxy is
burned to $^3He$ in protostellar  collapse. Galactic chemical evolution
models (Steigman and Tosi 1992, Vangioni-Flam, Olive and Prantzos 1994 [VFOP])
 show that $D/H$ can
be reduced in this way to its present interstellar value ($\approx
1.5\times 10^{-5}$,
Linsky  et al 1992) from any plausible initial
value. However, in the low mass
stars which now dominate the chemical recycling of the interstellar medium
(ISM),
the bulk of the material is assumed to be never heated to the higher
temperature
required to burn the $^3He$, so the bulk of
the primordial $D$ thus reappears
in the ISM as $^3He$ when the envelopes are ejected.
 For the galaxy as a whole, the sum
$(D+ ^3He)/H$ therefore only increases with time,  so that
this quantity can be used to set constraints on $(D/H)_p$.

But there are   empirical reasons for thinking that stellar populations
on average actually get rid of $^3He$.

The most widely used   measure of $^3He/H$ comes from measurements
of the solar wind, both from direct exposure experiments
and from meteorites, of   $^3He/ ^4He$ (Geiss 1993).
These are used to infer that  the   abundances of
the presolar nebula
  by number
were $D/H=2.6\pm 1.0\times 10^{-5}$,
$^3He/H=1.5\pm 0.3\times 10^{-5}$,
$(D+\ ^3He)/H=4.1\pm 1.0\times 10^{-5}$.
These are the most common numbers taken as Galactic or solar-circle
averages for the purpose of defining constraints on primordial abundances.

The only other useful  measure of cosmic $^3He/H$
 comes from  radio
emission maps  of highly ionized HII regions in the
Galaxy (Balser et al 1994, Wilson and Rood 1994).  The column density of
$^3He^+$ is
estimated from the brightness in the
 8.665 GHz hyperfine transition line, and the column
(squared) density of
$H^+$ or $^4He^+$ is estimated from   radio recombination lines.
Balser et al. use this data and a simple  model of the gas distribution
to obtain reliable estimates of $^3He/H$ in 7 Galactic HII regions,
and ``preliminary'' abundances and limits in 7 more.
Two of the most reliable ones are W43 and W49, with low values
$^3He/H= 1.13\pm 0.1 \times 10^{-5}$ and
 $^3He/H= 0.68\pm 0.15\times 10^{-5}$ respectively.
There appears to be a real  range of values, with W3 for
example measured at  $^3He/H= 4.22\pm 0.08\times 10^{-5}$,
and some are consistent with still higher values.
There may be a trend with galactocentric radius in the sense that
lower values tend to lie within the solar orbit and higher
values outside it.

Note that these results do not mesh with the standard picture.

1. The
Solar System value $(D+\ ^3He)/H$ is   greater than the interstellar one; if
$(D+\ ^3He)/H$ were  steadily increasing, it ought to be less, because of
the elapsed time since the formation of the solar system.

2. The
ISM shows large variations in $^3He/H$, which argues that one ought
not take any one point, such as the solar system, as an average
of the Galactic abundance.

3. The
gradient with Galactic radius goes the wrong way; if stars are creating
$^3He$ on average, it ought to be highly enriched towards the Galactic
center, like other heavier elements are.

4. If we adopt the lowest $^3He/H$ value in ISM (W49) as the primordial one,
assuming that the additional   $^3He$ found at other sites is Galactic in
origin
as required in the standard picture,
 the very low value requires a large $\eta\approx 2\times 10^{-9}$,
in which case SBBN predicts an excessive $^4He$ abundance
$Y_p\approx 0.26$. The   observed value is $Y_p=0.228\pm 0.005$
(Pagel et al 1992, see also Skillman and Kennicutt 1993), which is marginally
inconsistent even with the SBBN prediction for solar $^3He/H$, $Y_p=0.242$.

The
recent   determination of $D/H\approx 2.5\times 10^{-4}$
in a high redshift quasar absorber (Songaila et al 1994, Carswell et al
1994) highlights the
problems of interpreting the Galactic abundances. This estimate is
  more than an order of magnitude larger
than the interstellar $D/H$ values and a factor of 5 more than the solar system
estimates. If it is a real detection of deuterium it gives a firm upper
limit on $\eta$, since the big bang is the unique source of deuterium.
With the
small $\eta\approx 1.5\times 10^{-10}$
 implied by this measurement, SBBN works better, predicting close to  the
observed $^4He$
abundance ($Y_p=0.231$), and requiring a baryon density $\Omega h^2 = 0.005$
 close to that
found today in stars in gas, $\Omega h = 0.003$, so the
amount of baryonic dark matter is not large compared to that already seen.
Bearing in mind that this is still based on just one object which may be
contaminated by a
hydrogen interloper masquerading as deuterium, there is also no reason to
doubt this number,
except for the inconsistency with the standard interpretation of the
Galactic $^3He/H$
measurements.   In the long run  this type of measurement may be much more
reliable,
since the abundance is measured in pristine material, as demonstrated
by its very low observed metal abundances; at the very least, it allows a
sampling of
a wider variety of environments.

Even without this result,
the simplest interpretation of the
 empirical evidence in the Galaxy is that
 stellar populations, on average, tend to destroy not
only deuterium but also the $^3He$ that comes from it.

This simple picture has been elaborated by   quantitative
models of Galactic chemical evolution
(Vangioni-Flam, Olive and Prantzos 1994).
In these models, the fraction of initial $(D+ ^3He)/H$ returned is regarded
as a free parameter, or
as an adjustable function of stellar mass, $g_3$.
Motivated by the need to improve SBBN consistency,
VFOP showed that $(D+ ^3He)/H$  could be reduced in a model of galactic
chemical evolution,  but only if low-mass
stars are net destroyers of $^3He$ (i.e. $g_3<1$), in contradiction to
the usual assumptions of stellar models.
They were unable to identify a mechanism for this,
 but computed models anyway assuming
various modest destruction factors, and found models satisfying
a wide range of chemical constraints, including  a reduction of
$D/H$  by large factors, and matching the presolar $(D+ ^3He)/H$.
Thus, VFOP showed that
 a consistent picture is possible if $g_3$ is small for low mass stars,
but still lacked an explanation of how this can occur:
``we can offer
no solution as to why $g_3$ whould be lower other than
the constraints imposed by the presolar $D+ ^3He$ data.''

\bigskip
\vfil\eject

\section{$^3He$ Destruction in Low Mass Stars During Giant Branch Mixing  }

There are two reasons for supposing that Galactic stellar
populations are net producers of
$^3He$, but neither of them is airtight.

The first is that some sites have been found with really substantial
overabundances of $^3He$.
For example, a huge enhancement
$^3He/ ^4He\approx 0.7$ is   found (Hartoog 1979)
 in Feige 86; 3 Cen A is another example (Sargent and Jugaku 1961).
This may however  be purely an atmospheric effect
in rare   stars where
isotopic settling in thin photospheric layers leads to an
incorrect estimate of the true stellar composition (Vauclair and Vauclair
1982).
In another situation,
hyperfine emission in the  planetary nebula N3242
reveals (Rood, Bania and Wilson 1992)
 a large enrichment $^3He/H\approx 10^{-3}$, which seems
to demonstrate at least one local source.
However, it still could be that such enhancements are
rather rare, and that the generic behavior averaged over
 the stars that dominate
the recycling of the interstellar gas is destruction.

The second reason comes from  stellar evolution models, which
predict that low mass stars produce $^3He$ and eject it into
the ISM when they throw off their envelopes.  Although these models
are successful at predicting the shapes of HR diagrams,
temperatures, luminosities,  lifetimes, and principal nuclear burning products,
 they cannot
however be trusted when it comes to predicting the envelope abundances of
trace elements.
These elements can be changed significantly by
effects not in the models, which have little effect on the other
properties of the stars.

There is indeed
evidence that stars below about   $2\msol$ mix a substantial fraction of their
envelope material to
a nuclear burning zone at high temperature after they  leave the main sequence
for a long enough time to affect composition. The best evidence  comes
from high resolution spectroscopy of a variety of  main sequence and giant
branch stars in
Galactic open and globular clusters (Gilroy 1989; Gilroy and Brown 1991;
Brown, Wallerstein,
and Oke 1990; Sneden, Pilachowski and VandenBerg 1986). The abundances of
the  carbon
isotopes and of nitrogen are observed to follow a consistent pattern. The
models
correctly predict the
$^{12}C/ ^{13}C$ and $C/N$ ratios on the main sequence
and part way up the first giant branch, including ``first dredge up" of
processed material as the convection zone penetrates downwards to the core---
providing
reassuring confirmation   of the model predictions for the element profiles
right
down to the core. But near the top of the  giant branch the  ratios are
observed to fall well
below  the model predictions. There appears to be a concrete confirmation
of significant
alterations in the nuclear abundances of trace elements in the envelopes of
typical low mass
stars, which is what we are seeking.

The observed  ratios indicate that the material of the stars may have been
processed
to high enough temperatures,
for a long enough time,    for the elements to approach
$CN$ equilibrium. Note that
 the amount of burning time needed to change these
ratios appreciably is only one $CN$ cycle time, which corresponds roughly
to a fraction $C/H$ of the star's giant branch lifetime.
The data make sense if   all the
material in the envelope of the stars has been mixed down
to a hot burning zone, and each fluid element spends
spends   a total of at least one $CN$ cycle
time there after a star leaves the main sequence.

Charbonnnel (1994) offers an explanation of why this should happen in
low mass stars particularly. At low masses, the chemical composition
discontinuity left behind at the lowest point of contact reached by the
convective envelope
after dredge up, which inhibits mixing below this level by the steep gradient
in
molecular weight,  occurs inside the maximum radius achieved by the
hydrogen-burning shell
after the core becomes degenerate.  Thus for a time on the giant branch
the envelope can  mix
down to the hydrogen burning zone. The actual mechanism of the mixing is not
known. Although    candidate mechanisms have been suggested (Zahn 1992),
they are
not yet included in current stellar models (eg, Schaller et al 1992).

If this explanation of the observed
$^{12}C/ ^{13}C$ and $C/N$ ratios is correct,
the same process would likely have a strong influence on $^3He$.
The observed reduction of $^{12}C/ ^{13}C$ to nearly equilibrium values
implies that the bulk of the  material in the envelope  reaches
a high enough temperature long enough for
thorough conversion; as we see below, the  high temperatures are
 more than adequate to also burn
the $^3He$ in the time available.  A similar argument applies
for the $C/N$ ratios.

We can use the data from one of Charbonnel's models
(a $1.25\msol$, $Z=0.02$ star) to estimate the timescales and
temperatures involved.   The data imply that
conversion begins after   the onset of mixing; for this
model, the shell reaches the chemical discontinuity, and
mixing can start, at step 33 of
Schaller et al's table 19, at  $5.43\times 10^9$y.
The conversion must be complete   well before
 the star leaves the giant branch at the helium flash (step 51,
$5.55\times 10^9$y); thus the conversion of envelope material must take
place in
substantially less than $12\times 10^7$y.
{}From  Charbonnel's  figure 9, the  mass of  the   shell where the
$^{12}C\rightarrow\ {^{13}C}$ process occurs is only  about 0.001 of  the
total mass of the
star.
By continuity, since the entire envelope is
processed in the conversion zone then the cumulative time spent by any mass
element in
the
zone is less than 0.001 of the total available time, or about
 $10^5$y. But for any reasonable density, for the
$^{12}C\rightarrow {^{13}C}$ reaction to occur in $10^6$y or $10^5$y
the temperature reached  must  be at least
$T_6>15$ or $T_6>17$ respectively, where $T_6=T/10^6 K$
(see e.g. Clayton 1968). (Note that the central
temperature at  the onset of mixing is $T_6=34$.)

Although we cannot observe the $^3He$ in these giants directly,
we can make a quantitative estimate of the  effect
on $^3He$ by comparing its destruction rate with that of
$^{12}C$.
The dominant destructive
interaction will be $^3He(\alpha,\gamma)^7Be$ (rather than
$^3He(^3He,2p)^4He$ which dominates for abundances $\ge 10^{-4}$).
The rate-limiting reaction of the $CN$ processing  is
$^{12}C(p,\gamma)^{13}N$, which we already know must be important---
that is, a typical $C$ nucleus must undergo at least of the order
of one reaction. Assuming a ratio of $\alpha$ to $p$ of 0.1 by number,
and using    reaction rates
 from Fillipone (1986) and Caughlan and Fowler (1988), we estimate  the
ratio of the $^{12}C$-destruction time to the $^3He$-destruction time,
independent of other abundances or density (e.g. Clayton 1968):
$$
R\equiv {\tau_{12}\over\tau_3}=\exp[{14.3 T_6^{-1/3}- 2.5}].
$$
At  the required  $T_6=15$, $R=27$; for $T_6=17$, $R=21$.
Thus over the likely range of maximum temperatures
reached by the envelope material, the near equilibration of the
$^{12}C\rightarrow ^{13}C$
 implies equilibration of  $^3He\rightarrow ^4He$; in fact the
$^3He$ will be destroyed more quickly, and could occur even in stars which
do not show the  $^{12}C/ ^{13}C$ anomaly.

Since the $^3He\rightarrow ^4He$ rates are faster than the $CN$ rates,
which has apparently approached equilibrium
at high temperature, the local
 abundance  $^3He/ ^4He$ in the high temperature region must also have
reached
 the
equilibrium value. For
$T_6>15$, the equilibrium abundance is   $^3He/H<10^{-5}$.
Thus the $CN$ data imply that the bulk of the envelope material must have been
processed through conditions where  $^3He/H<10^{-5}$.

It is not clear from these simple arguments
 that this is the abundance of the material when
it returns to the upper envelope, for it must pass through
lower-temperature regions to get
there, where its abundance might increase again. The actual returned
abundance   depends
on the detailed temperature history of a fluid element.
But quantitatively, to maintain net destruction requires only   that the
upward flow pass
through the region where
$T_6\approx 10$ in less than about $10^6$y, the time for regenerating
the $^3He$, which is certainly plausible.
Recall  that  each fluid element spends
an integrated total of
less than $10^5$y lingering in the hotter  $T_6\approx 15$ zone,
which would be an {\it upper limit} for the flow timescale in
a ``typical'' process where the material turns over more than once.
(Note that if the upward flows are
 concentrated in narrow plumes or updrafts, the fluid velocity
 is higher going up and the time spent
 traversing the region where $^3He/H$ increases is even shorter.)
Whether or not the material reaches the upper envelope without
also having $ ^3He/H$ increase significantly can only be determined
in the context of a detailed stellar model including both the mixing flow
and the nuclear reactions, but it would not be surprising   if
red giant branch mixing reduces
  the initial   $(D+\ ^3He)/H$ in the ejected envelope by a large enough factor
 ($g_3<0.1$) to solve the empirical difficulties discussed above.

The conclusion is not   that   $^3He$ must be destroyed,
but that there are good reasons to suspect that it might be, and
that therefore  Galactic $(D+\ ^3He)/H$
should not be used as an indicator of
or constraint on primordial $D/H$.

\acknowledgments

I am grateful for many useful discussions with D. Brownlee,
J. Brown, G. Fuller, and  G. Wallerstein.
The work was supported by NASA grants NAGW-2569 and NAGW-2523 at the
University of Washington.


\begin{references}


\reference Balser, D. S., Bania, T. M., Brockway, C. J., Rood, R. T., and
Wilson,
T. L., 1994, \apj, in press

\reference Brown, J. A., Wallerstein, G. W., and Oke, J.B. 1990, \aj, 100,
1561, 1990

\reference Carswell, R. F., Rauch, M., Weymann, R. J.,  Cooke, A. J.,
and Webb, J. K., 1994, MNRAS, 268, L1-L4

\reference Caughlan,G. R., and Fowler, W. A., 1988, Atomic Data Nucl. Data
Tables, 40, 283

\reference Charbonnel, C. 1994, AA, 282,811


\reference Clayton D.D. 1968,``Principles of Stellar Evolution and
Nucleosynthesis'',
New York: McGraw-Hill


\reference Copi, C. J., Schramm, D. N., and Turner, M. S., 1994, preprint
FERMILAB-Pub-94/174-A, (sissa: astro-ph 9407006)

\reference Fillipone, B. W. 1986, Ann. Rev. Nucl. Part. Sci., 36, 715

\reference Geiss, J. 1993, in Origin and Evolution of the Elements, ed. N.
Prantzos,
E. Vangioni-Flam and M. Cass\'e (Cambridge: Cambridge Univ. Press), p. 89

\reference Gilroy, K.K., 1989, \apj, 347, 835

\reference Gilroy, K. K. and Brown, J. A., 1991, \apj, 371,578


\reference Hartoog, M. 1979, \apj, 231,161

\reference Linsky, J. L., et al., 1992, \apj, 402, 694

\reference  Pagel, B. E. J., Simonson, E. A., Terlevich, R. J., \&
Edmunds, M. G., 1992, MNRAS, 255, 325


\reference Rood, R. T., Bania, T. M., and   Wilson, T. L. 1992, Nature,
355, 618

\reference Sargent, W. L. W., and Jugaku, J., 1961, \apj, 134, 777

\reference Schaller, G., Schaerer, D., Meynet, G., and Maeder, A. 1992, AA
Suppl, 96,269

\reference Skillman, E. D., and Kennicutt, R. C., 1993, \apj, 411, 655

\reference Smith, M. K.,  Kawano, L.H., and Malaney, R.A., 1993, \apj Supp,
85,219

\reference Sneden, C., Pilachowski, C. A., and VandenBerg, D. A., 1986,
\apj, 311, 826
\reference Songaila, A., Cowie, L.L., Hogan, C. J., and Rugers,
M. 1994, Nature, 368, 599

\reference Steigman, G., and Tosi, M. 1992, \apj, 401, 150

\reference Vangioni-Flam, E., Olive, K. A., and Prantzos, N., 1994, \apj,
427,618

\reference Vauclair, G., and Vauclair, S., 1982, Ann. Rev. A. Ap., 20, 37

\reference Wilson, T. L., and Rood, R. T., 1994, Ann. Rev. A. Ap., 32, in press

\reference Zahn, J. P., 1992, AA, 265, 115

\end{references}
\end{document}